Nanoscale Imaging of Lithium Ion Distribution During *In Situ* Operation of Battery Electrode and Electrolyte

*Megan E. Holtz,[1+] Yingchao Yu,[2+] Deniz Gunceler,[3] Jie Gao,[2] Ravishankar Sundararaman,[3] Kathleen A. Schwarz,[2] Tomás A. Arias,[3] Héctor D. Abruña,[2] David A. Muller[1,4*]*

[1] School of Applied and Engineering Physics, Cornell University, Ithaca, NY 14853

[2] Department of Chemistry and Chemical Biology, Cornell University, Ithaca, NY 14853

[3] Department of Physics, Cornell University, Ithaca, NY 14853

[4] Kavli Institute at Cornell for Nanoscale Science, Cornell University, Ithaca, NY 14853

[+] These authors contributed equally to this work.

*Corresponding author

**A major challenge in the development of new battery materials is understanding their fundamental mechanisms of operation and degradation. Their microscopically inhomogeneous nature calls for characterization tools that provide *operando* and localized information from individual grains and particles. Here we describe an approach that images the nanoscale distribution of ions during electrochemical charging of a battery in a transmission electron microscope liquid flow cell. We use valence energy-loss spectroscopy to track both solvated and intercalated ions, with electronic structure fingerprints of the solvated ions identified using an *ab initio* non-linear response theory. Equipped with the new electrochemical cell holder, nanoscale spectroscopy and theory, we have been able to determine the lithiation state of a $LiFePO_4$ electrode and surrounding aqueous electrolyte in real time with nanoscale resolution during electrochemical charge and discharge. We follow lithium transfer between electrode and electrolyte and observe charging dynamics in the cathode that differ among individual particles. This technique represents a general approach for the *operando* nanoscale imaging of electrochemically active ions in a wide range of electrical energy storage systems.**

The integration of renewable, and often intermittent, energy sources such as solar and wind into the energy landscape, as well as the electrification of transportation, requires dramatic advances in electrical energy conversion and storage technologies including fuel cells, batteries and supercapacitors.[1,2] Advancing our understanding necessitates the development of experimental tools capable of *operando* characterization that can discern mechanisms of operation and degradation in the native operating environment. Energy storage materials, such as battery electrodes, often display inhomogeneous behavior on the nanoscale.[3] Thus, the most illuminating and useful characterization methods are those capable of providing detailed mechanistic information of charge/discharge dynamics of *individual* grains and particles. TEM investigations specialize in revealing structural and compositional information with nanoscale spatial resolution and sub-second temporal resolution. Unfortunately, conventional TEM is not compatible with studies of many processes related to electrochemical energy storage because they take place in liquid environments. Recently, the development of TEM holders that encapsulate thin liquid layers promise *in situ* imaging and spectroscopy on the nanoscale.[4-8] Incorporating electrodes[9,10] enables *in situ* imaging of electrochemical processes,[11-13] electrodeposition[9] and dendrite growth.[14] However, quantitative electrochemistry in the microscope remains a major challenge: standard silicon fabrication techniques introduce electrochemically active species into the environment, and unconventional electrode shapes and configurations may lead to species migration, large background currents, and large uncompensated resistances. Here we develop broadly-applicable, *quantitative* electrochemistry in a liquid cell TEM holder that can be correlated with microstructure and local electronic structure changes during operation, even for surface-sensitive catalysts such as those used in fuel cells. To follow the underlying ion redistributions, we demonstrate a method for spectroscopic imaging of nanoscale processes during electrochemical operation and follow the charging and discharging dynamics of a battery electrode.

The lithium-ion battery is a particularly promising candidate for electric vehicle and energy storage applications.[1,15-18] A key mechanism in the performance of lithium-ion battery electrodes is how the lithium ions intercalate and deintercalate from the electrode during cycling. Here, as a demonstration of tracking lithiation and degradation in an *in situ* battery in the TEM, we studied the cathode material LiFePO$_4$, which has surged in interest due to its attractive capacity, ability to sustain high charging and discharging rates, abundance, low toxicity, relative operational safety, and low cost.[17,19] There is much discussion in the literature on the mechanism of lithiation and delithiation,[3,17,20-27] with evidence of a two-phase reaction or a metastable solid solution. Within the two-phase reaction pathway, there are different theories for the propagation of lithiation. Some of the disagreement may be attributed to many-particle effects, where bulk measurements (both *in situ* and *ex situ*) convolve signals from the entire area of the electrode probed.[3] *Ex situ* studies are inherently compromised by removal of the particles from their native – and often reactive – environment, which leads to questions of relaxation or reactions caused by the foreign surroundings. Here we use a liquid cell *in situ* TEM, which can probe, in real time, the evolution of individual grains and nanoparticles in the native environment of a battery.

TEM detection of lithium through a liquid is difficult, because lithium is a weak elastic scatterer and multiple scattering from the liquid overwhelms the inelastic core-loss signal in electron energy-loss spectroscopy (EELS). In this work, we successfully observed the lithiation state by valence energy-filtered TEM (EFTEM), which probes the low-energy regime (~1-10 eV), and allows us to work in thicker liquid layers than core-level spectroscopy.[6] Valence EELS can provide electronic structure information, allowing us to track battery charging and discharging as ions are being transferred between electrode and electrolyte. This method is analogous to observing color changes in optical spectroscopy during battery cycling on the micron scale,[28,29] except that valence EFTEM achieves nanometer resolution. While electronic structure features in the lithiated and delithiated electrode are often well documented,[30] identifying electronic structure fingerprints in the solution are less well explored because there are many solvated species in solution that are difficult to distinguish. Here we employed *ab initio* theory to calculate optical gaps of solvated species.[31] We took solution effects into account with a hybrid functional[32] including a nonlinear description of the polarization response of the surrounding liquid,[33] which gives a more physical model of bound solvent charges near the solute than linear models. For the first time, we applied this technique to calculate excited electronic state and found quantitative matches to experimental excitation energies. By combining electrochemistry in the TEM with valence spectroscopic imaging and theory, we were able to identify the lithiation state of the electrode *and* electrolyte during *in situ* operation.

**A Baseline for Electrochemistry in the TEM**

We use a liquid cell holder developed by Protochips using chips we designed to mimic a typical electrochemical cell (Figure 1a-b). The tip of the holder is a microfluidic flow cell with silicon nitride viewing membranes that confine a liquid, shown in cross section in Figure 1a. Figure 1b illustrates the top chip, with three patterned electrodes optimized for electrochemical cycling and imaging. Traditional silicon-processing methods use a chromium adhesion layer and gold electrodes. However, chromium diffuses rapidly through gold (especially at grain boundaries) and can affect and even dominate the electrochemical signal. In addition, high-atomic-number electrodes such as Pt and Au obscure imaging. Instead, we used a carbon working electrode which only weakly scatters electrons and is commonly used in bulk electrochemistry, and titanium adhesion layers under platinum reference and counter electrodes. This allows us to image through the electrode with little loss in spatial resolution and contrast, which is dominated by scattering in the liquid instead. As a practical matter, and discussed below, spatial resolution is often limited by the low doses needed to control radiation damage than by beam spreading in the cell.

To demonstrate that *in situ* electrochemistry reproduces well-established criteria, we performed cyclic voltammetry of a film of platinum, shown in Figure 1c-d, in the TEM. This control experiment represents a test case for quantitative electrochemistry, since the features are surface effects – including hydrogen adsorption and desorption and oxide formation and reduction – which are sensitive to contaminants at the sub-monolayer level. The *in situ* electrochemistry reproduced the characteristic voltametric profile of a polycrystalline platinum electrode at an appropriate current scale, regardless of the electron beam. In thin liquid layers, the ohmic drop in the solution becomes significant, as evidenced by the slanted curve in Figure 1d. This implies an inherent compromise between the highest spatial resolution imaging and quantitative electrochemistry. Accounting for ohmic drops in solution, this setup replicates results of a conventional electrochemical cell while obtaining nanometer resolution.

Having established the electrochemical performance of the TEM holder, we studied LiFePO$_4$, a widely used Li-ion cathode material, in 0.5 M Li$_2$SO$_4$ aqueous electrolyte. The *ex situ* characterization of the LiFePO$_4$ is discussed in Supplementary Materials Section 1. Because aqueous electrolyte shows safety benefits over carbonates, and due to its high abundance, low weight and non-toxicity, researchers have considered aqueous electrolytes in addition to the more traditional carbonates.[34] We find aqueous electrolyte is practical for technique development: carbonates are more viscous than aqueous electrolyte, leading to higher flow pressures and potentially more window breakages. In the event of electrolyte leakage, aqueous liquid will dissipate quickly while carbonate electrolytes lead to contamination of the microscope column.

**Spectroscopy and *Ab Initio* Studies to Determine Lithiation State**

To elucidate the lithiation mechanism we must examine how the lithium ions intercalate into the electrode. In the TEM, lithiation can be tracked by morphological changes or structural changes using electron diffraction,[21,35] although morphology does not give chemical maps and diffraction spots are quickly obscured in thicker liquid films. As a light element, lithium scatters electrons weakly, making elastic imaging challenging, and the energy-dispersive x-ray signal for lithium has too low an energy to detect. We instead tracked the lithiation state of the battery using EELS, which offers chemical fingerprints (core-loss EELS) and electronic structure information (valence EELS). Yet, using EELS to identify lithium in liquids has two obstacles. First, EELS is degraded by multiple scattering events in thick liquids.[6] Second, the lithium-K edge resides at 54 eV and is lost in the superimposed bulk plasmon of the thick films of liquid. Additionally, the lithium-K edge overlaps with many transition metal M edges such as iron.[36] This makes core-loss EELS of the lithium practically impossible to detect in the liquid cell TEM.

Valence EELS, which interrogates electronic structure, can detect the state of lithiation of battery electrodes in liquid electrolytes. During discharge and charge, lithium ions move in and out of the electrode, filling and emptying valence bands, thereby changing the electronic structure. These electronic structure shifts are accessible by optical spectroscopy, where lithiation has been observed on the micron scale in electrodes as they change color.[28,29] Valence EELS surveys the same electronic levels as ultraviolet-visible (UV-VIS) spectroscopy. Optical absorption spectra track the imaginary part of the energy-dependent dielectric constant, $Im(\varepsilon)$, and the electron energy-loss function in EELS is proportional to $Im(\varepsilon) / [Re(\varepsilon)^2 + Im(\varepsilon)^2]$. An advantage of valence EELS is high spatial resolution, which is ultimately limited to the nanoscale by the delocalization of the low-energy excitations.[37] While delocalization prevents atomic-resolution valence EFTEM studies, resolution in the liquid environment is often more strongly limited by multiple scattering or by the low-dose imaging conditions required. Valence EELS provides strong signals due to large scattering cross sections and low background from the liquid. The electronic structure shift usually occurs where the electrolyte is stable, at energies below its optical gap (~6-7 eV) where the electrolyte is transparent. Harnessing the electronic structure shifts in battery electrodes during cycling is a practical method to track the lithiation state.

The spectroscopic characteristics of the battery cathode LiFePO$_4$ and electrolyte 0.5 M Li$_2$SO$_4$/H$_2$O are shown in Figure 2. The monochromated valence EELS of dry LiFePO$_4$ is shown in Figure 2a. There is a fingerprint of the delithiated FePO$_4$ at 5 eV, which is not present in the lithiated LiFePO$_4$. This corresponds to the electronic structure shift in FePO$_4$ as it lithiates to LiFePO$_4$. As lithium ions interact with the Fe-3d bands, the corresponding peak at 5 eV disappears.[30,38] This peak enables quick spectroscopic mapping of the state of lithiation.[39]

Figure 2b presents the UV-VIS spectra of 0.5 M Li$_2$SO$_4$ electrolyte, and for comparison 0.5 M H$_2$SO$_4$ and water. There is a peak in the 0.5 M Li$_2$SO$_4$ solution at 6.2 eV, not present in sulfuric acid or water. Because of the high pH in the Li$_2$SO$_4$ electrolyte, there is a very low concentration of protonated species. Using equilibrium constants we can identify the solvated species in solution: 0.74 M Li$^+$, 0.26 M LiSO$_4^-$, 0.24 M SO$_4^{2-}$, with less than $10^{-3}$ M of LiOH, HSO$_4^-$ and LiHSO$_4$. We uniquely identified that the absorption peak at 6.2 eV in the electrolyte is due to solvated LiSO$_4^-$ using *ab initio* theory. Because electronic screening from the surrounding electrolyte shifts the optical gaps of the relevant species on the same order as their separations (~1 eV), we developed a novel *ab initio* approach to calculate excited states while accounting for the surrounding liquid. We employed joint density-functional theory (JDFT) to compute the electronic structure information of solvated species in thermodynamic equilibrium with a liquid environment.[31] Using a hybrid functional for the solute[32] and a nonlinear description of the polarization response of the surrounding liquid,[33] this approach yielded an *ab initio* optical gap of solvated LiSO$_4^-$ of 6.3 eV, close the experimental absorption peak at 6.2 eV. Thus, the presence of this peak indicates a lithiated solution. Other solvated species in the solution have peaks at higher energies, which are discussed and tabulated in Supplementary Material Section 2. This theory shows remarkable agreement with experiment, and has been repeated for other solvated ions measured by UV-VIS, see Supplementary Materials Section 2. In the 5 eV EFTEM images we have two fingerprints: one of the delithiated cathode material $FePO_4$ at 5 eV, and one for the lithiated liquid electrolyte at 6.2 eV.

We used EFTEM to obtain spectroscopic mapping of the 5 eV signal with a 5 eV wide slit (2.5 to 7.5 eV), which captured the state of lithiation of both the particle and the solution. The 0 eV and 5 eV spectroscopic images of $LiFePO_4$ particles in a 200 nm thick 0.5 M $Li_2SO_4$ electrolyte are shown in Figure 2c-d, respectively. In the 0 eV EFTEM image, the particles appear fairly homogeneous. In the 5 eV EFTEM image, the delithiated regions of the particles are brighter, enabling us to differentiate delithiated and lithiated particles rapidly on the nanoscale. The solution has a high intensity, indicating a lithiated solution – as expected in equilibrium. We used electron beam conditions that minimized beam interactions in a control experiment, where irradiation without cycling had no apparent affect on morphology or composition (Supplementary Figure 3). Another control experiment with the same electron beam conditions showed similar effects over the entire electrode, not just in the location imaged (Supplementary Figure 4). This method of EFTEM enables quick (second-long) mapping of the nanoscale lithium distribution in electrode and electrolyte.

**Tracking Lithiation State of Electrode and Electrolyte During *In Situ* Cycling**

Having an electrochemical cell for the TEM and a technique to observe the lithiation state, we assembled an *in situ* battery using an activated carbon anode, 0.5 M $Li_2SO_4$ aqueous electrolyte, and a $LiFePO_4$ cathode. We imaged at 5 eV with a 5 eV wide energy window to track the state of lithiation (Figure 3a, Supplementary Materials Movie 1) and recorded electrochemical data (Figure 3b-c) simultaneously. Figure 3b shows the galvanostatic charge/discharge experiment with ±10 nA current applied between the anode and cathode. From an estimate of the amount of active material present on the electrode and the specific capacity from *ex situ* aqueous studies (31 mAh/g, Supplementary Figure 1), this corresponds roughly to a charge/discharge rate of about 10 C (10 cycles per hour). Figure 3c shows the resulting voltage profile between the anode and cathode. Because the potential difference for the deintercalation (intercalation) of lithium ions between $LiFePO_4$ and $FePO_4$ is 1 V,[40] charging (discharging) occurs in the potential range of our experiment. The rapid cycling rate enabled multiple charge-discharge cycles to be acquired in the course of the experiment and decreased the electron beam exposure time.

There are clear differences in the 5 eV spectroscopic images between the charged (Figure 3a, right) and the discharged state (Figure 3a, left) in both the particles and the solution. In the charged state, compared to the discharged state, particles show more bright regions - indicated by white arrows - corresponding to delithiated $FePO_4$. Additionally, the cluster of particles is overall brighter in the charged image, especially around the edges of the cluster, than in the discharged image, as marked by black arrows. The brightest particles may correspond to completely transitioned $FePO_4$, whereas the overall slight increase in intensity in the particles may indicate partially delithiated particles. On discharge, these bright regions of $FePO_4$ disappear, transitioning back to $LiFePO_4$. If we spatially integrate the 5 eV EFTEM intensity over the particle, shown in red in Figure 3d, we see an increase in intensity on charge and a decrease on discharge, compared to the solution far away from the particles (black). The intensity of the particles in Figure 3d was raised to the background level of the solution. Radiation damage is expected to be irreversible and uncorrelated with voltage cycle, and the appearance of the bright regions of $FePO_4$ and lithiated solution is repeatable and correlated with charge state – indicating the electron beam did not cause the signals observed. This demonstrates tracking of the lithiation state of battery electrodes at the nanoscale.

We next examined the 5 eV EFTEM intensity in the solution adjacent to the particles. There is a local decrease in intensity in the solution surrounding the particles during discharge, which can be seen in Figure 3a. The spatially integrated signal from the solution adjacent to the particle (blue) drops dramatically during discharge, plotted in Figure 3d. From UV-VIS measurements and JDFT calculations, the bright intensity in the solution is caused by $LiSO_4^-$. As the particles lithiate during discharge, the adjacent solution becomes depleted of $Li^+$ and $LiSO_4^-$, causing the drop in the 5 eV signal. The profile of the intensity drop matches that of a diffusional concentration profile (Supplementary Figure 2), supporting that it is due to depletion of species near the electrode. Additionally, the intensity change appears in the inelastic but not in the elastic images, indicating a chemical change. We thus observe the expected delithiation of the solution in the 5 eV EFTEM images as the particles are being lithiated. Thus valence EFTEM can track the lithium ions in the particles and solution during battery cycling.

**Inhomogeneity and Lithiation Mechanisms**

With the capability to locate ions at the nanoscale, we explored the mechanism of lithiation and delithiation of individual cathode nanoparticles. There are several proposed mechanisms of lithiation for $LiFePO_4$,[21,23-25,41] which have been reported to depend on particle size, coating, synthesis methods, charging rate and experimental conditions.[42] These methods typically rely on bulk particle analysis which convolutes many particle effects.[3] We observed the evolution of many individual particles under high rate conditions in aqueous solution. The evolution of one cluster of particles is shown in Figure 4, corresponding to the voltage profile in Figure 4a. In Figure 4b the cell is discharged, and the particles and solution are dark, corresponding to a lithiated particle and a delithiated solution. During charge, the orange arrows track the evolution of a particle. In Figure 4c, the start of nucleation is seen. In Figure 4d, we see a core-shell type structure, which completely transforms into $FePO_4$ in Figure 4e. In Figure 4f, the particle appears to have mostly fractured off. We track the evolution of another representative particle denoted by the yellow arrows, where the edge of the particle transitions to $FePO_4$, and the delithiation front propagates anisotropically across the particle until it is completely delithiated in Figure 4g. We return to the discharged state in Figure 4h, and the bright regions disappear, converting to $LiFePO_4$.

13solution in the 5 eV EFTEM images as the particles are being lithiated. Thus valence EFTEM can track the lithium ions in the particles and solution during battery cycling.

**Inhomogeneity and Lithiation Mechanisms**

With the capability to locate ions at the nanoscale, we explored the mechanism of lithiation and delithiation of individual cathode nanoparticles. There are several proposed mechanisms of lithiation for $LiFePO_4$,[21,23-25,41] which have been reported to depend on particle size, coating, synthesis methods, charging rate and experimental conditions.[42] These methods typically rely on bulk particle analysis which convolutes many particle effects.[3] We observed the evolution of many individual particles under high rate conditions in aqueous solution. The evolution of one cluster of particles is shown in Figure 4, corresponding to the voltage profile in Figure 4a. In Figure 4b the cell is discharged, and the particles and solution are dark, corresponding to a lithiated particle and a delithiated solution. During charge, the orange arrows track the evolution of a particle. In Figure 4c, the start of nucleation is seen. In Figure 4d, we see a core-shell type structure, which completely transforms into $FePO_4$ in Figure 4e. In Figure 4f, the particle appears to have mostly fractured off. We track the evolution of another representative particle denoted by the yellow arrows, where the edge of the particle transitions to $FePO_4$, and the delithiation front propagates anisotropically across the particle until it is completely delithiated in Figure 4g. We return to the discharged state in Figure 4h, and the bright regions disappear, converting to $LiFePO_4$.

The delithiation of individual particles as seen in Figure 3a and in Figure 4 demonstrates slow nucleation during the transformation. Growth of the phase is also slow enough for us to image (Figure 4) and we see particles that are not fully transformed – in contrast with the "domino cascade" model that predicts the rapid growth and full transformation of individual particles once nucleated. We observed core-shell structures, but more commonly delithiation started at an edge and then moved through the rest of the particle supporting anisotropic growth. Also, stronger regions of delithiation are seen on the edges of agglomerates, where the particle may be in better electrical contact with the current collector. However, the same particles are not always the active ones. The kinetics are consistent with a diffusional response (Supplementary Figure 3), which is not surprising, considering the high cycling rate and thin liquid layer. Strikingly, the particles exhibit an inhomogeneous response at the nanoscale and many of the particles are inactive at any moment in time. This inhomogeneous response is likely a characteristic of the kinetics and the mechanism of Li-ion insertion and de-insertion associated with multi-particle polycrystalline $LiFePO_4$. This highlights the advantages of nanoscale imaging during cycling, as bulk analysis can rarely deconvolve these effects.

We observe degradation mechanisms in the $LiFePO_4$ particles during the course the rapid charge/discharge cycles. We see gradual mass loss of the $LiFePO_4$ throughout the experiment from our observations from elastic 0 eV EFTEM images (20% particle area reduction in 5 cycles). The fracturing and mass loss were observed in a control experiment to occur over the entire electrode, even where it was not exposed to the electron beam during cycling (Supplementary Figure 4). This is consistent with our observations in the 5 eV EFTEM images during cycling. As particles delithiated, they often disappeared from the field of view, followed by a formation of another delithiated region – seen in Figure 4e and 4f. An explanation is that as the particles delithiate, and given the extreme cycling conditions, lattice strain causes regions to fracture away from the particle. In fact, fracturing has been observed in *ex situ* studies.[20,43]. After fracturing, a fresh surface of $LiFePO_4$ is exposed, enabling further delithiation.

This work demonstrates the unique ability of liquid cell *in situ* TEM coupled with spectroscopy and theory to observe the lithiation insertion and de-insertion dynamics and degradation of $LiFePO_4$ in real time. These techniques may provide valuable insights into operation and other degradation pathways in a wide range of electrical energy conversion and storage devices such as batteries, fuel cells and supercapacitors.

**Methods**

We used a Protochips *in situ* holder. The liquid flow cell portion of the design has been discussed previously.[44] We flowed electrolyte at 100-300 μL/hr to ensure no depletion of species or accumulation of electron beam damaged solution. The new addition of three electrodes (Figure 1a,b) in the microfluidic cell enables electrochemical studies under well defined conditions. On the viewing membrane, we deposited the material of interest onto the carbon working electrode that scatters electrons weakly, facilitating imaging on the electrode. The platinum reference electrode is close to the working electrode to minimize uncompensated resistances – although the cell potential in the battery experiments were measured between counter and working electrodes. The platinum counter electrode is large to provide current, and far away from the working electrode to minimize material migration to the working electrode. The chips were prepared prior to use as previously discussed.[6]

Imaging and EELS were performed using a monochromated FEI Tecnai F-20 STEM/TEM operated at 200 kV and equipped with a Gatan 865 HR-GIF spectrometer for EELS analysis. With the monochromator filter, the energy resolution was 0.2 eV. Even when the monochromator was not employed, the study did benefit from the system's improved energy stability with an energy resolution of 0.6 eV. For the data acquisition that resulted in Figures 3 and 4, the 5 eV EFTEM image was continuously recorded using a 2 s long acquisition time with periodic elastic imaging at 0 eV to observe overall morphology. Electron beam conditions were carefully selected, (500 e-/nm$^2$s) so as not to cause changes to the particles without cycling (see Supplementary Figure 3).

$LiFePO_4$ was synthesized by a solid-state reaction.[45] Specifically, a mixture of $Li_2CO_3$, $FeC_2O_4 \cdot 2H_2O$ and $NH_4H_2PO_4$ (molar ratio: Li/Fe/P=1/0.9/0.95) was ball-milled for 2 hours under argon using a Spex8000 mixer. The ball-milled precursor was first heated at 350 °C for 10 hours under argon. After cooling down, the precursor was ground in a mortar and then heated at 600 °C for 10 hours under argon to get the final product, which was tested *ex situ* as described in the Supplementary Material 1. The $LiFePO_4$ nanoparticles were roughly 100-200 nm in diameter. They were dispersed in an ethylene glycol and isopropyl alcohol solution and were printed onto the working electrode of the electrochemical cell chip using a Dimatix printer. The mass of nanoparticles deposited was estimated to be 30 ng from the solution concentration and the drop size. An excess amount of activated carbon was applied to the counter electrode (anode).

A Gamry potentiostat was used. We were able to achieve normal electrochemical processes, which would occur in a typical microelectrode experiment, in the holder in the microscope. All optical ultraviolet-visible (UV/VIS) absorption spectra were obtained using a HP 8453 diode array spectrometer at room temperature in the denoted solvents, with a conventional 1.0 cm quartz cell.

To compute optical gaps within JDFT, we considered the bound charges from the electrolyte to be fixed during the excitation because by far the greatest screening effects in aqueous electrolytes are due to nuclear rearrangements (either motion of ions in the fluid or reorientation of water molecules), which occur over much longer time scales than such electronic excitations. To ensure that the potential provided by these bound charges remained fixed during the excitation, we computed HOMO-LUMO gaps of the relevant ions in the potential associated with the *ground state* of the relevant species. All *ab initio* calculations employed our open-source density-functional software JDFTx,[31] with the relevant species and their first solvation shells treated explicitly using density-functional theory at the PBE0 exchange-correlation functional level[32], and the liquid treated at the level of a nonlinear polarizable continuum.[33] Values from these studies are shown in Supplementary Table 1, showing remarkable agreement with the observed UV-VIS level, and now affording unique identification of the origin of each level.

**Acknowledgements**

We would like to thank John Grazul and Mick Thomas for assistance in the TEM facility, John Damiano and David Nackahashi at Protochips, and Pinshane Huang and Huolin Xin for valuable comments and discussions. This work made use of the electron microscopy facility of the Cornell Center for Materials Research (CCMR) with support from the National Science Foundation Materials Research Science and Engineering Centers (MRSEC) program (DMR 1120296). This work and MEH and YY are supported by the Energy Materials Center at Cornell, an Energy Frontier Research Center funded by the U.S. Department of Energy, Office of Basic Energy Sciences under Award Number DESC0001086. YY also acknowledges the fellowship from the American Chemical Society Division of Analytical Chemistry sponsored by Eastman Chemical Co.

**Author Contributions**

DAM, HDA, YY and MEH designed the electrochemical TEM LFP experiment. MEH and YY performed the experiments. MEH analyzed the data and wrote the paper. DG and RS, supervised by TA, designed and performed JDFT calculations to analyze the solution intensity, with discussions with KAS. JG synthesized the LFP particles.

**Competing Financial Interests Statement**

The authors declare no competing financial interests.

**References**

1	Tarascon, J. M. & Armand, M. Issues and challenges facing rechargeable lithium batteries. *Nature* **414**, 359-367, doi:10.1038/35104644 (2001).

2	Whittingham, M. S. History, Evolution, and Future Status of Energy Storage. *Proceedings of the Ieee* **100**, 1518-1534, doi:10.1109/jproc.2012.2190170 (2012).

3	Malik, R., Abdellahi, A. & Ceder, G. A Critical Review of the Li Insertion Mechanisms in LiFePO4 Electrodes. *Journal of the Electrochemical Society* **160**, A3179-A3197, doi:10.1149/2.029305jes (2013).

4	de Jonge, N. & Ross, F. M. Electron microscopy of specimens in liquid. *Nat. Nanotechnol.* **6**, 695-704, doi:10.1038/nnano.2011.161 (2011).

5	Yuk, J. M. *et al.* High-Resolution EM of Colloidal Nanocrystal Growth Using Graphene Liquid Cells. *Science* **336**, 61-64, doi:10.1126/science.1217654 (2012).

6	Holtz, M. E., Yu, Y., Gao, J., Abruña, H. D. & Muller, D. A. In Situ Electron Energy-Loss Spectroscopy in Liquids. *Microsc. microanal.* **19**, 1027-1035, doi:doi:10.1017/S1431927613001505 (2013).

7	Jungjohann, K. L., Evans, J. E., Aguiar, J. A., Arslan, I. & Browning, N. D. Atomic-Scale Imaging and Spectroscopy for In Situ Liquid Scanning Transmission Electron Microscopy. *Microsc. microanal.* **18**, 621-627, doi:10.1017/s1431927612000104 (2012).

8       Zheng, H. M. *et al.* Observation of Single Colloidal Platinum Nanocrystal Growth Trajectories. *Science* **324**, 1309-1312, doi:10.1126/science.1172104 (2009).

9       Williamson, M. J., Tromp, R. M., Vereecken, P. M., Hull, R. & Ross, F. M. Dynamic microscopy of nanoscale cluster growth at the solid-liquid interface. *Nat. Mater.* **2**, 532-536, doi:10.1038/nmat944 (2003).

10      Grogan, J. M. & Bau, H. H. The Nanoaquarium: A Platform for In Situ Transmission Electron Microscopy in Liquid Media. *J. Microelectromech. Syst.* **19**, 885-894, doi:10.1109/jmems.2010.2051321 (2010).

11      Unocic, R. R. *et al. In-situ* Characterization of Electrochemical Processes in Energy Storage Systems. *Microsc. microanal.* **17**, 1564-1565 (2011).

12      Huang, J. Y. *et al.* In Situ Observation of the Electrochemical Lithiation of a Single SnO2 Nanowire Electrode. *Science* **330**, 1515-1520, doi:10.1126/science.1195628 (2010).

13      Liu, X. H. *et al.* Ultrafast Electrochemical Lithiation of Individual Si Nanowire Anodes. *Nano Letters* **11**, 2251-2258, doi:10.1021/nl200412p (2011).

14      White, E. R. *et al.* In situ transmission electron microscopy of lead dendrites and lead ions in aqueous solution. *ACS Nano* **6**, 6308-6317 (2012).

15      Dahn, J. R., Zheng, T., Liu, Y. H. & Xue, J. S. MECHANISMS FOR LITHIUM INSERTION IN CARBONACEOUS MATERIALS. *Science* **270**, 590-593, doi:10.1126/science.270.5236.590 (1995).

16   Whittingham, M. S. ELECTRICAL ENERGY-STORAGE AND INTERCALATION CHEMISTRY. *Science* **192**, 1126-1127, doi:10.1126/science.192.4244.1126 (1976).

17   Padhi, A. K., Nanjundaswamy, K. S. & Goodenough, J. B. Phospho-olivines as positive-electrode materials for rechargeable lithium batteries. *Journal of the Electrochemical Society* **144**, 1188-1194, doi:10.1149/1.1837571 (1997).

18   Herle, P. S., Ellis, B., Coombs, N. & Nazar, L. F. Nano-network electronic conduction in iron and nickel olivine phosphates. *Nat. Mater.* **3**, 147-152, doi:10.1038/nmat1063 (2004).

19   Chen, J. J. & Whittingham, M. S. Hydrothermal synthesis of lithium iron phosphate. *Electrochemistry Communications* **8**, 855-858, doi:10.1016/j.elecom.2006.03.021 (2006).

20   Chen, G. Y., Song, X. Y. & Richardson, T. J. Electron microscopy study of the LiFePO4 to FePO4 phase transition. *Electrochem. Solid State Lett.* **9**, A295-A298, doi:10.1149/1.2192695 (2006).

21   Brunetti, G. *et al.* Confirmation of the Domino-Cascade Model by LiFePO4/FePO4 Precession Electron Diffraction. *Chemistry of Materials* **23**, 4515-4524, doi:10.1021/cm201783z (2011).

22   Jones, J. L., Hung, J. T. & Meng, Y. S. Intermittent X-ray diffraction study of kinetics of delithiation in nano-scale LiFePO4. *J. Power Sources* **189**, 702-705, doi:10.1016/j.jpowsour.2008.08.055 (2009).

23  Laffont, L. *et al.* Study of the LiFePO4/FePO4 two-phase system by high-resolution electron energy loss spectroscopy. *Chemistry of Materials* **18**, 5520-5529, doi:10.1021/cm0617182 (2006).

24  Delmas, C., Maccario, M., Croguennec, L., Le Cras, F. & Weill, F. Lithium deintercalation in LiFePO(4) nanoparticles via a domino-cascade model. *Nat. Mater.* **7**, 665-671, doi:10.1038/nmat2230 (2008).

25  Sharma, N. *et al.* Direct Evidence of Concurrent Solid-Solution and Two-Phase Reactions and the Nonequilibrium Structural Evolution of LiFePO4. *J. Am. Chem. Soc.* **134**, 7867-7873, doi:10.1021/ja301187u (2012).

26  Malik, R., Zhou, F. & Ceder, G. Kinetics of non-equilibrium lithium incorporation in LiFePO4. *Nat. Mater.* **10**, 587-590, doi:10.1038/nmat3065 (2011).

27  Dreyer, W. *et al.* The thermodynamic origin of hysteresis in insertion batteries. *Nat. Mater.* **9**, 448-453, doi:10.1038/nmat2730 (2010).

28  Harris, S. J., Timmons, A., Baker, D. R. & Monroe, C. Direct in situ measurements of Li transport in Li-ion battery negative electrodes. *Chem. Phys. Lett.* **485**, 265-274, doi:10.1016/j.cplett.2009.12.033 (2010).

29  Patel, M. U. M. *et al.* Li-S Battery Analyzed by UV/Vis in Operando Mode. *Chemsuschem* **6**, 1177-1181, doi:10.1002/cssc.201300142 (2013).

30  Kinyanjui, M. K. *et al.* Origin of valence and core excitations in LiFePO4 and FePO4. *J. Phys.-Condes. Matter* **22**, 275501, doi:275501 10.1088/0953-8984/22/27/275501 (2010).

31   Sundararaman, R., Letchworth-Weaver, K. & Arias, T. A. *JDFTx*, (2012). Available from http://jdftx.sourceforge.net.

32   Ernzerhof, M. & Scuseria, G. E. Assessment of the Perdew-Burke-Ernzerhof exchange-correlation functional. *J. Chem. Phys.* **110**, 5029-5036, doi:10.1063/1.478401 (1999).

33   Gunceler, D., Letchworth-Weaver, K., Sundararaman, R., Schwarz, K. A. & Arias, T. A. The importance of nonlinear fluid response in joint density-functional theory studies of battery systems. *Modelling and Simulation in Materials Science and Engineering* **21**, 074005 (2013).

34   Wang, Y. G., Yi, J. & Xia, Y. Y. Recent Progress in Aqueous Lithium-Ion Batteries. *Adv. Energy Mater.* **2**, 830-840, doi:10.1002/aenm.201200065 (2012).

35   Wang, C.-M. *et al.* In Situ Transmission Electron Microscopy Observation of Microstructure and Phase Evolution in a SnO2 Nanowire during Lithium Intercalation. *Nano Letters* **11**, 1874-1880, doi:10.1021/nl200272n (2011).

36   Moreau, P. & Boucher, F. Revisiting lithium K and iron M-2,M-3 edge superimposition: The case of lithium battery material LiFePO4. *Micron* **43**, 16-21, doi:10.1016/j.micron.2011.05.008 (2012).

37   Muller, D. A. & Silcox, J. DELOCALIZATION IN INELASTIC-SCATTERING. *Ultramicroscopy* **59**, 195-213, doi:10.1016/0304-3991(95)00029-z (1995).

38   Sigle, W., Amin, R., Weichert, K., van Aken, P. A. & Maier, J. Delithiation Study of LiFePO4 Crystals Using Electron Energy-Loss Spectroscopy. *Electrochem. Solid State Lett.* **12**, A151-A154, doi:10.1149/1.3131726 (2009).

39	Moreau, P., Mauchamp, V., Pailloux, F. & Boucher, F. Fast determination of phases in LixFePO4 using low losses in electron energy-loss spectroscopy. *Appl. Phys. Lett.* **94**, 123111, doi:10.1063/1.3109777 (2009).

40	Luo, J. Y., Cui, W. J., He, P. & Xia, Y. Y. Raising the cycling stability of aqueous lithium-ion batteries by eliminating oxygen in the electrolyte. *Nat. Chem.* **2**, 760-765, doi:10.1038/nchem.763 (2010).

41	Ramana, C. V., Mauger, A., Gendron, F., Julien, C. M. & Zaghib, K. Study of the Li-insertion/extraction process in LiFePO4/FePO4. *J. Power Sources* **187**, 555-564, doi:10.1016/j.jpowsour.2008.11.042 (2009).

42	Zhang, W. J. Structure and performance of LiFePO4 cathode materials: A review. *J. Power Sources* **196**, 2962-2970, doi:10.1016/j.jpowsour.2010.11.113 (2011).

43	Wang, D. Y., Wu, X. D., Wang, Z. X. & Chen, L. Q. Cracking causing cyclic instability of LiFePO4 cathode material. *J. Power Sources* **140**, 125-128, doi:10.1016/j.jpowsour.2004.06.059 (2005).

44	Klein, K. L., Anderson, I. M. & De Jonge, N. Transmission electron microscopy with a liquid flow cell. *J. Microsc.* **242**, 117-123, doi:10.1111/j.1365-2818.2010.03484.x (2011).

45	Kang, B. & Ceder, G. Battery materials for ultrafast charging and discharging. *Nature* **458**, 190-193, doi:10.1038/nature07853 (2009).

**Figures**

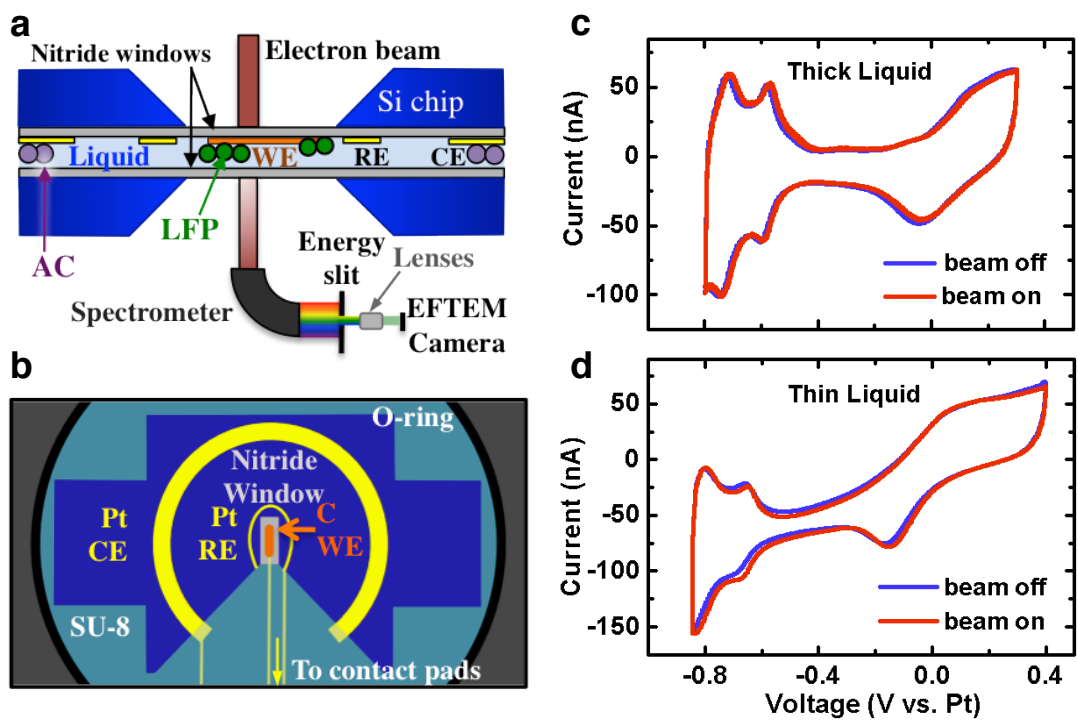

**Figure 1.** Schematic of the *in situ* electrochemistry holder and electrochemical data. (a) Cross-sectional view of the holder, with silicon nitride membranes encapsulating a fluid layer. The working electrode (WE), made of carbon, lies in the viewing window, with LiFePO$_4$ (LFP) nanoparticles deposited on top. The platinum counter electrode (CE) is coated with an excess of activated carbon (AC). In EFTEM mode, energies are selected by a slit to be imaged. (b) Schematic of the top chip, with three patterned electrodes: a carbon WE on the viewing membrane, Pt RE (not used here) and Pt CE. The connection leads are covered by SU8, and the contact pads to the holder do not contact the liquid, so as to minimize electrochemical activity outside the viewing window. The chips exhibited electrochemical activity qualitatively similar to that of an *ex situ* microelectrode, as shown for the Pt cyclic voltammetry (CV) in (c) and in (d). In extremely thin liquid layers (~150 nm) the voltammetric profile exhibits a significant ohmic drops as seen in (d).

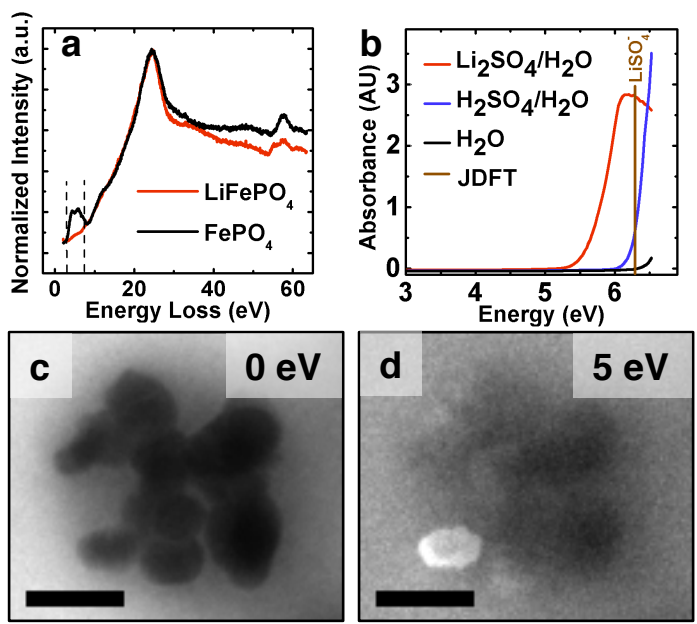

**Figure 2.** Spectroscopy of LiFePO$_4$ and the delithiated counterpart FePO$_4$ and the aqueous electrolyte Li$_2$SO$_4$. (a) Monochromated EELS with an energy resolution of 0.2 eV of a dry sample shows a 5 eV peak for FePO$_4$ but not LiFePO$_4$. (b) UV-VIS spectra of the electrolyte 0.5 M Li$_2$SO$_4$/H$_2$O, 0.5 M H$_2$SO$_4$/H$_2$O, and pure water. There is an absorption peak at 6 eV for the Li$_2$SO$_4$ solution. The JDFT calculated gaps of the solvated species in solution reveal the 6 eV peak is caused by LiSO$_4^-$. EFTEM of LiFePO$_4$ in 0.5 M Li$_2$SO$_4$/H$_2$O with a 5 eV energy slit around (c) 0 eV where the liquid dominates the signal and the particles look fairly homogeneous and at (d) 5 eV, which highlights the FePO$_4$. Scale bar is 200 nm. Using the 5 eV EFTEM image we can locate delithiated regions.

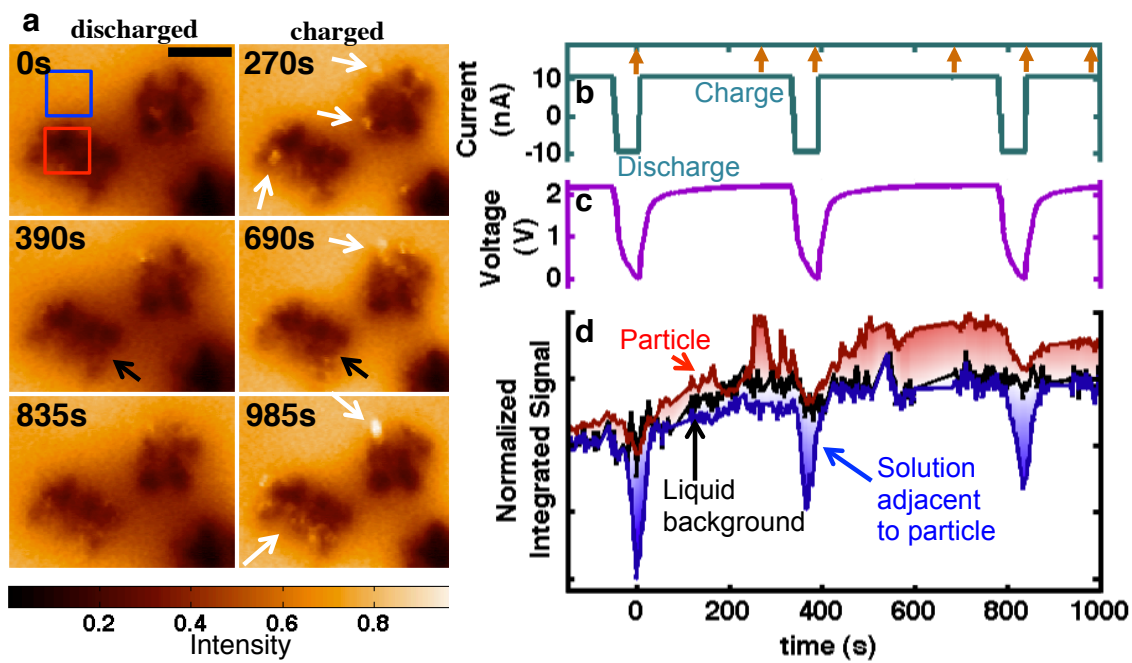

**Figure 3.** Charging and discharging of the cathode material LiFePO$_4$ *in situ* in 0.5 M Li$_2$SO$_4$ aqueous electrolyte. (a) 5eV spectroscopic EFTEM images of charging and discharging are shown with a 400 nm scale bar, corresponding to times marked in (c). Bright regions are delithiated FePO$_4$ and dark regions are LiFePO$_4$. There are more bright regions of FePO$_4$ at the end of charge cycles and less during the discharges. (b) Current profile corresponding to 10C. The corresponding voltage profile is in (c), referencing the activated carbon counter electrode. (d) Integrated intensity over various regions, tracking with the voltage profile, from the regions shown by the boxes in (a5). Line profiles across the particle corresponding to highlighted region in (a5) as a function of time are plotted in (e). The solution becomes very dark during discharges and returns to the background level during charge. Regions of the particle are seen to light up and disappear, potentially due to delithiating and fracturing off of the particle cluster. During times when no imaging occurred, the data are linearly extrapolated, and for comparison the intensity is brought to the same level by subtraction.

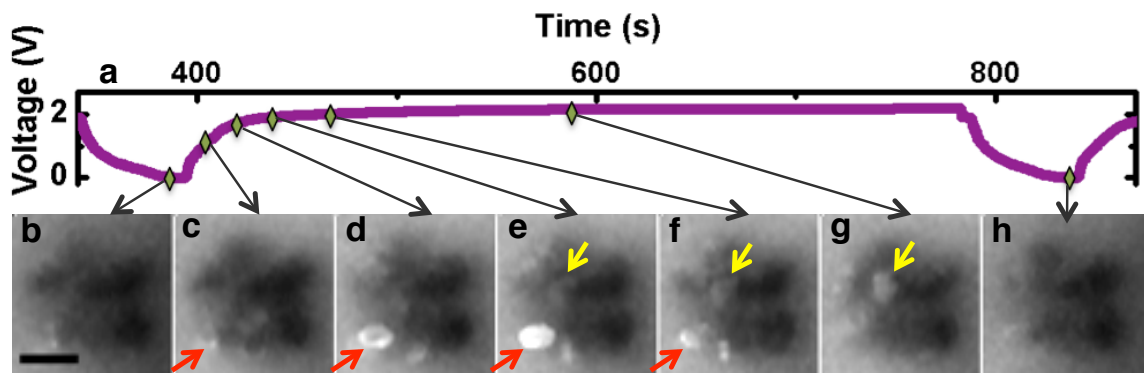

**Figure 4.** Observing the evolution of a cluster of LiFePO$_4$/FePO$_4$ during one charge/discharge cycle. The voltage profile of the second cycle is shown in (a). The 5 eV EFTEM image in (b) is completely discharged, with a scale bar of 200 nm. At the bottom of (c) and (d) we see the starting of core-shell structures. In (d) a bright particle appears with a core-shell structure which fills in brighter in (e), which partially disappears in (f). More regions of bright FP develop in (g), and the particle returns to discharged in (h) where it is darker. In general, the charged images (d-g) have more bright regions than the images taken in the discharged state, (a) and (h), which have significantly more dark sections. Arrows indicate particles delithiating by a core-shell pathway (red) and one starting from one edge propagating through the particle (yellow).